# Carbon nanotubes for coherent spintronic devices

F. Kuemmeth*, H. O. H. Churchill, P. K. Herring, C. M. Marcus

*Department of Physics, Harvard University, Cambridge, Massachusetts 02138, USA*

*email: kuemmeth@physics.harvard.edu

Abstract:

Carbon nanotubes bridge the molecular and crystalline quantum worlds, and their extraordinary electronic, mechanical and optical properties have attracted enormous attention from a broad scientific community. We review the basic principles of fabricating spin-electronic devices based on individual, electrically-gated carbon nanotubes, and present experimental efforts to understand their electronic and nuclear spin degrees of freedom, which in the future may enable quantum applications.

Carbon nanotubes (NTs) have been studied by material scientists since 1952[1], but became a worldwide research focus only after fullerenes were discovered and after single-walled carbon nanotubes (SWNTs) were introduced to the research community in 1993[2]. The simplicity of their synthesis and the diversity of their properties[3] quickly propelled NTs into electronics, optics, nano- and biotechnology research labs around the globe. Today's nanofabrication techniques allow access to individual nanotubes, enabling novel integrated circuits[4,5], fast[6] or flexible[7] transistors, nanomechanical oscillators[8-10] and photoactive devices[11-13].

While the basic electronic properties of NTs have been the subject of previous reviews[14,15], new quantum mechanical effects have been discovered recently. These are related to confinement of carriers in a quantum dot (i.e. a short NT segment displaying discrete energy levels), and to spin (i.e. intrinsic angular momentum and magnetic moment), and may shape the design of future quantum technologies. In this review we describe methods of fabrication of nanoscale quantum dot (QD) devices based on individual NTs, and discuss how they provide an understanding of electronic and nuclear spins and their interactions in NTs. These efforts are a step toward spintronic devices[16] and solid-state quantum computation[17] based on NTs.

After reviewing the basics of NT synthesis and electronic properties we focus on three recent experiments: The fabrication of clean, suspended QDs and their role in revealing spin-orbit interactions[18], the fabrication of top-gated, highly tunable double quantum dots (DQDs) and their response to nuclear spins[19], and the potential of charge sensing and pulsed-gate techniques to study electronic spin dynamics[20].

# Carbon nanotube synthesis

The synthesis of NTs is simple and inexpensive. For basic research, where small quantities of high quality SWNTs are needed, chemical vapor deposition (CVD) is the preferred method: gaseous carbon compounds are decomposed in a furnace at temperatures around 1000C and nanotubes are nucleated from nanometer scale catalyst particles[21, 22]. For the experiments described below, either methane or ethylene gas, combined with hydrogen, was used. Resulting SWNTs had diameters of $d \sim$ 1-2 nm.

**Isotopic engineering**

Nanotubes synthesized from natural hydrocarbons consist of $\sim$ 99% $^{12}$C and $\sim$ 1% $^{13}$C. Changing this isotopic composition can be important for several reasons. First, the mass difference between $^{12}$C and $^{13}$C directly affects phonon modes in the nanotube. By changing the $^{13}$C concentration of the growth gas during NT synthesis and using spatially resolved Raman spectroscopy, L. Liu *et al.*[23] elucidated NT growth mechanisms (Fig. 1a, b). Due to their distinct Raman frequencies, isotopically controlled and DNA-functionalized NTs can also serve as selective and bright biomarkers, as recently demonstrated by H. Dai and coworkers[24]. Second, $^{13}$C possesses a nuclear spin of 1/2 (in units of $\hbar$, the quantum of angular momentum), while $^{12}$C has zero nuclear spin. Electron-nuclear spin interactions in GaAs-based spin qubits have been a fascinating research area[25-27], motivating new spin-based experiments using NTs. By annealing SWNTs filled with $^{13}$C enriched fullerenes, F. Simon *et al.*[28] created double-wall nanotubes with an inner tube that was predominantly $^{13}$C, identified by the redshift of the Raman radial breathing mode of the tube (Fig. 1c). Such structures allow, for instance, nuclear magnetic resonance studies of $^{13}$C SWNTs that are shielded from each other by $^{12}$C shells. Isotopically engineered NT structures may one day allow quantum devices that use electron spins for the manipulation and readout of quantum information and nuclear spins for its storage[29]. Developments in this direction will require a detailed understanding of both spin-orbit coupling as well as electron-nuclear (hyperfine) coupling in nanotubes. This long-term challenge is a principal motivation for the experiments described in this review.

**Nanotube characterization**

Independent of isotopic composition, NTs grown under identical conditions will exhibit diverse electronic properties due to uncontrolled variation in diameters and chirality. A wide range of device parameters and interesting quantum phenomena are encountered, ranging from massive, strongly interacting quasiparticles[30, 31] to massless fermions that travel unimpeded by scattering from Coulomb potentials[32, 33].

The devices we describe here all employ electric potentials produced by gate electrodes to confine and manipulate the tunneling of individual carriers. Because the band gap determines the effective mass of a carrier, this parameter is useful for engineering tunneling rates. Typically, a back gate (a conductive back plane of the chip on which the sample is grown) is used to assess overall gating properties of NTs at room temperature. Gate response allows the band gap to be estimated, but provides no information about chirality. Techniques under development, based on microscopic photoluminescence imaging[34,35], light scattering[36,37], as well as mechanical transfer methods[38] will allow fabrication of optical, mechanical, and electronic NT devices with known chirality.

## Electronic properties

**Small band gap nanotubes**

Tight-binding models suggest that NTs are either semiconducting, with a band gap inversely proportional to diameter ($E_{gap} \sim 0.7$ eV·nm/$d$), or metallic, with a linear dispersion, $E = \hbar v_F k$, where $v_F \sim 8*10^5$ m/s is the Fermi velocity[39] and $k$ is the wave vector, depending on chirality[3, 40, 41]. In practice, most metallic nanotubes possess small band gaps[42-44] of tens of meV, presumably due to curvature[45], strain[46, 47] and electron-electron interactions[48]. While semiconducting nanotubes are attractive for their optical properties and room-temperature electronic devices, small band gap nanotubes display small effective masses ($E_{gap}=2mv_F^2$) and are therefore ideal for many QD experiments in which tunnel couplings depend on both the barrier potential and the effective mass. Light mass also mitigates the effects of the disorder, which is present in all devices. However, since the barrier potential itself cannot exceed the band gap due to Klein tunneling[33], the band gap should be sufficiently large to prevent unwanted barrier transparency via Klein tunneling.

**Valley degeneracy and large orbital moments**

It is instructive to visualize the electronic structure of NTs in terms of the linear dispersion of graphene, which occurs at two points in momentum space (K and K') due to the inversion symmetry of the graphene lattice. Taken from these Dirac cones, the only allowed states for a NT are those that comply with the quantization condition of fitting an integer number of Fermi wavelengths around the circumference of the NT (Fig. 2a). When the closest quantization line (green) misses the K point, a band gap appears along with hyperbolic electron-like and hole-like dispersions near the K point. Ignoring spin for the moment, time-reversal symmetry guarantees a second set of energy bands with exactly the same energy at the K' point. For a confining potential that is smooth on the atomic scale, discrete quantum states can be formed from either the K or K' valley, yielding a two-fold degenerate energy spectrum. The valley degeneracy constitutes a discrete, two-state quantum degree of freedom (termed isospin) that is insensitive to long-wavelength electrical noise, and so is potentially useful as a long-lived quantum two-level system, or qubit. Isospin, combined with spin, gives a four-fold degeneracy in the electronic spectrum.

One implication of the Dirac-cone picture is that stationary states formed from one valley (green dot in Fig. 2a) carry a persistent current around the nanotube circumference, while the opposite valley (purple dot) carries the opposite current. Magnetic moments associated with these "clockwise" and "counterclockwise" currents are remarkably large, equivalent to several Bohr magnetons ($\mu_{orb} \approx 3.4 \mu_B \cdot d$/nm), and therefore couple strongly to external magnetic fields applied parallel to the nanotube axis[49] (Fig. 2b). Ex-ternal fields can thus be used to tune both the band gap and the energy separation of opposite valley states (Fig. 2c).

## Suspended carbon nanotubes

**Merits of suspension**

Probing the intrinsic symmetries of NTs is challenging unless care is taken during nanofabrication to minimize randomizing factors such as disorder from the substrate. Non-suspended NTs have revealed the four-fold degeneracy associated with valley and spin symmetry, manifested in a four-fold shell structure[50, 51] in weakly coupled QDs, as well as in a SU(4) Kondo effect[52] in the strongly correlated regime. Other phenomena, such as electron-hole symmetric level spacings, have only been realized in suspended NTs[53]. The cleanest possible transport studies are achieved in suspended NT devices if no device processing is required *after* CVD growth[54]. Besides circumventing disorder from the substrate and/or due to organic residue from fabrication, these devices allow transport studies in a regime where electrons interact strongly due to the large "effective fine structure constant" of nanotubes: $e^2/(2\varepsilon_0 h v_F) \sim 1$. Due to their versatile applications and scientific value for the study of exotic quantum phenomena like Mott insulation[48] and Wigner crystallization[31], we briefly describe the fabrication of clean, suspended NT devices. We mention that progress has been made to develop suspended, tunable DQDs with few electrons and holes[33], and that suspended NTs also allow their exceptional mechanical properties to be used for nanoscale radio-frequency signal processing, ultra-sensitive mass detection[55], and high-Q[56] non-linear[9, 10] resonators[8].

**Fabrication**

Critical to realizing clean, suspended NT devices is that electrical contacts and gates be compatible with high temperatures and chemistry of CVD nanotube growth. Devices shown in Fig. 3 were fabricated by etching and thermally oxidizing doped silicon on an insulating oxide, resulting in electrically isolated gate electrodes (Fig. 3a). After patterning contacts based on metals with high melting points (such as W or Pt, Fig. 3c) and dispensing catalyst particles[21], the chip was loaded into the CVD furnace. Due to the random growth of nanotubes, only a small fraction of devices contains a single, suspended nanotube with good electrical contact and appropriate band gap. It is therefore necessary to fabricate large arrays of potential devices with various trench sizes on each chip (Fig. 3d) and to characterize their conductance and gating properties after NT growth. Fig. 3b shows 4 out of 108 devices on a 6 mm x 6 mm silicon chip (the coloring is due to the thickness variation of the polished silicon layer). Using a vacuum probe station at various temperatures (room temperature, 70 K and 4 K), the most promising devices are selected for wire bonding and measurements in a dilution refrigerator.

**Spin-orbit coupling**

In the absence of disorder, electron-electron interactions, and spin-orbit coupling, the ground state of the one-electron NT QD is four-fold degenerate, reflecting both spin (↑/↓) and valley (K/K') degeneracies (Fig. 4a). To resolve the four states experimentally via tunneling spectroscopy, a magnetic field parallel to the nanotube axis of 300 mT was applied to couple to the spin and orbital moments of a QD containing a single electron[18]. This results in four parallel features in Fig. 4b whose distances are proportional to the energy differences between the four quantum states. Moreover, their magnetic field dependence (Fig. 4c) reveals that there are indeed two "clockwise orbits" (K↑ and K↓) and two "counterclockwise orbits" (K'↑ and K'↓). Surprisingly, not all four states become degenerate at B=0. The electron states with parallel orbital- and spin magnetic moment (K↓ and K'↑) appear slightly lower in energy than the states with antiparallel alignment (K↑ and K'↓), while the opposite is observed for a one-hole QD. These findings are explained in terms of spin-orbit coupling, and indicate that such effects cannot be neglected in very clean NTs. Theory suggests that this type of spin-orbit coupling – first predicted by T. Ando a decade ago[57] and now an active theoretical research field[58-67] – may allow spin-manipulation by electrical means[68] as well as optical control of quantum information[69-71].

## Top-gated quantum dots

### Fabrication

Device fabrication using electron-beam lithography and atomic layer deposition of gate oxides yields highly-tunable DQDs with integrated charge sensors. These devices allow independent control of charge states and tunnel barriers and constitute a powerful platform to study the electron-nuclear (hyperfine) interaction and spin coherence in $^{13}$C and $^{12}$C nanotubes. Fabrication proceeds as follows: Pt/Au alignment marks are patterned by electron-beam lithography followed by patterning of an array of 5 nm thick Fe catalyst pads on a small chip (~ 5 mm on a side) of degenerately doped thermally oxidized silicon. The chip is then loaded into a CVD furnace (Fig. 5b) that uses either $^{12}$C or $^{13}$C methane feedstock. Upon identifying straight nanotube segments using a scanning electron microscope (Fig. 5a), devices are contacted with Pd patterned using electron beam lithography and metal lift-off. Devices are then coated with a 30 nm $Al_2O_3$ top-gate insulator using atomic layer deposition (ALD). To preserve the electronic properties of the NTs, a non-covalent functionalization layer (Fig. 6a) using iterated exposure to $NO_2$ and trimethylaluminum[72] is applied before the $Al_2O_3$ ALD process (Fig. 6b). The high dielectric constant of $Al_2O_3$ enhances the capacitive coupling of the NT to aluminum electrodes (top gates) (Fig. 5f).

### Conductance through a double quantum dot

The gate electrodes (shaded blue in Fig. 5f) create three tunnel barriers: one each to the source and drain contacts and one between the left and right dots. Each quantum dot can hold between 0 and hundreds of electrons or holes depending on the combination of voltages applied to the barrier and plunger gates (shaded green).

The conductance through the DQD sensitively depends on the voltages applied to the gate electrodes. The continuous Coulomb oscillations in Fig. 7a (measured at -1 mV source-drain bias and temperature ~ 100 mK) indicate that at these gate voltages, the DQD merged into a single QD, coupled approximately equally to the left and right gate electrodes. By making the gate voltages more negative (i.e. repelling electrons) and raising the middle barrier (i.e. by also setting $V_M$ more negative), the number of charges ($N_L$, $N_R$) within each QD becomes quantized, with Coulomb blockade operating in both dots. This suppresses the overall conductance and limits current flow to specific combinations of gate voltages ("triple points") that simultaneously lift Coulomb blockade in both dots, giving the hexagonal "honeycomb pattern" seen in Fig. 7b. Transport through NT DQDs has been studied by various groups[73-77] with a particular emphasis on spin physics[19, 77, 78].

## Electron-nuclear interactions

### Pauli blockade

At finite source-drain bias, energy conservation allows current to flow through the DQD only in triangular regions near each triple point[79] in gate-voltage space. Current near the base of the triangle (dashed line in Fig. 8a) reflects the probability of tunneling from the ground state of one dot into the ground state of the other. If electron spin played no role, the current would be symmetric in applied bias. The experimental data, however, is not symmetric in bias, as seen in Fig. 8b. This asymmetry, familiar in semiconductor quantum dots and known as spin or Pauli blockade, arises from the filling of degenerate orbital levels for asymmetrically loaded devices. In NTs, selection rules for both spin and isospin can lead to a generalized Pauli blockade[80]. An example of Pauli blockade, involving spin but not isospin, is shown for illustrative purposes in Figs. 8c,d: For forward bias, whenever an electron tunnels into the singlet ground state of (0,2) it can tunnel into the left dot (1,1) without changing its spin. For reverse bias, however, whenever an electron happens to tunnel into a triplet state in (1,1), it cannot tunnel into the (0,2) singlet ground state without flipping its spin (in (0,2) the triplet state is energetically inaccessible due to the Pauli exclusion principle). If spin relaxation is slow, these blocking events suppress the average reverse base current, making Pauli blockade a sensitive probe of spin dynamics. When both spin and isospin are involved, Pauli blockade is more complex, but operates in a similar way.

### Hyperfine coupling

Hyperfine interaction between confined electrons and the large number ($\sim 10^5$) of thermally randomized nuclear spins in $^{13}$C QDs are expected to affect both the spin and isospin lifetime[81]. However, due to the conservation of energy and the different Zeeman splitting of electrons and nuclei, hyperfine-mediated lifting of Pauli blockade should only be observable near zero magnetic field. Indeed, the reverse base current in the $^{13}$C device of Fig. 8b shows a sharp maximum at B=0 (Fig. 8e), which is not seen in the $^{12}$C devices[19]. If one estimates the strength of the hyperfine coupling from the width of this low-field feature, one obtains an estimate for the hyperfine coupling that is two orders of magnitude higher than expected from the small admixture of s-orbitals due to the nanotube's curvature[82]. We note that in $^{12}$C devices with stronger interdot tunneling the reverse base current displays a minimum at B=0 (Fig. 8e, bottom panel), suggesting that broken time-reversal symmetry facilitates relaxation through spin-orbit coupling. Similar behavior has been observed in InAs nanowires[83].

## Pulsed-gate double quantum dots

**Charge sensing**

The ability to measure charge states rather than conductance greatly simplifies DQD spin-qubit readout. As demonstrated in nearly isolated GaAs DQDs, qubit states can be coherently manipulated by electrical pulses applied to top gates on a nanosecond time scale[84], and charge states can be read out within a few microseconds by measuring the conductance of a quantum point contact[85]. Pauli blockade converts spin information into charge information, and enables the study of spin relaxation[26] and dephasing[84] times by charge sensing techniques. The charge state of nanotube QDs can be measured by monitoring the conductance through a separately contacted single electron transistor fabricated from oxidized aluminum near the nanotube device[86,87], or from the same nanotube (Fig. 5g). In the latter case it is capacitively coupled to the DQD with an electrically floating wire (orange). Tuned to the edge of a Coulomb oscillation using gate electrodes, the sensor conductance is sensitive to the potential of the coupling wire and hence to the charge configuration of the DQD. Using charge sensing techniques, pulsed spectroscopy of nearly isolated NT QDs and real time detection of tunneling rates as low as 1 Hz have been demonstrated[87].

**Pulse triangles**

Figure 9 shows the conductance of the charge sensor, $g_s$, as a function of two gate voltages applied to the DQD. Without electrical pulses (Fig. 9a) the sensor conductance shows four distinct plateaus, corresponding to (0,1), (1,1), (1,2), or (0,2) occupation of the DQD. When appropriate electrical pulses are added to the left and right gate voltage (and repeated cyclically) the average sensor conductance shows a fifth value inside the white triangle in Fig. 9b. This conductance value lies between the (1,1) and (0,2) values and is the signature of Pauli blockade. In particular, position M displays the average sensor conductance while the DQD is repeatedly emptied (50 ns spent at E), reset into a (1,1) state with random spin and isospin orientations (R, 50 ns), and pulsed to the measurement point (M, 500 ns). The ground state near M is (0,2), but only some of the states loaded at R can tunnel to the (0,2) configuration due to Pauli blockade. Therefore, the sensor conductance is closer to the (1,1) plateau than it would be in the absence of Pauli blockade. If the waiting time at the measurement point, $\tau_M$, is increased, more of the blocked (1,1) states are able to relax and tunnel into (0,2), and the sensor conductance approaches the (0,2) plateau. The dependence of the sensor conductance on $\tau_M$ therefore reveals the relaxation time of the DQD (Fig. 9c). In this device[20] we find that the relaxation time decreases with increasing magnetic field, possibly due to phonon-mediated spin relaxation enabled by spin-orbit coupling, a mechanism that is suppressed at small magnetic fields by Van Vleck cancellation[88].

# Outlook and future challenges

We have described fabrication of controllable quantum devices based on carbon nanotubes. This system provides a versatile platform for studying spin and valley selection rules and relaxation, as well as hyperfine coupling in $^{13}$C devices. Given the variability of performance observed between different devices, it will be important to distinguish effects arising from intrinsic variations (such as chirality) and extrinsic variations (disorder due to nanofabrication processes). While suspended devices allow optical identification[37] of their chirality, as well as clean, tunable DQDs[33], they currently do not allow a straightforward integration of charge sensing, a powerful tool for the readout of qubit states.

We note that SWNTs grown from purified $^{12}$C offer an alternative to other group-IV materials which are being pursued as nuclear-spin free host materials for quantum applications[89-92]. Future work on $^{12}$C nanotubes includes studying the predicted chirality and diameter dependence[62] of spin-orbit coupling and demonstrating that single spins can be manipulated coherently using time-dependent electric fields. Measurements of dephasing times in purified $^{12}$C devices will be a step toward solid state qubits free of decoherence due to nuclei.

Major open questions for $^{13}$C are whether the surprisingly large hyperfine coupling inferred from transport experiments reflects a correct interpretation of the data, what role is played by anisotropic hyperfine coupling expected from p-orbitals, and how nuclear polarizations affect Pauli blockade and relaxation. Unlike in GaAs QDs, in which millions of nuclei of the QD are constantly interacting with a macroscopic ensemble from the GaAs substrate, the nuclei of a $^{13}$C nanotube may constitute a useful, fully coherent entity rather than an incoherent thermal bath.

In summary, the future of NT-based spintronic applications depends on a thorough understanding and clever use of the electronic properties related to spin, isospin and isotopic composition.

# Acknowledgements

We thank Max Lemme and Shahal Ilani for helpful discussions, and acknowledge support through the National Science Foundation, ARO/iARPA, the Department of Defense, Harvard's and Cornell's Center for Nanoscale systems, and the Cornell node of the National Nanotechnology Infrastructure Network.

# Bibliography

1. Radushkevich, L. V.; Lukyanovich, V. M., *Zurn. Fisic. Chim.* (1952) **26**, 88
2. Iijima, S.; Ichihashi, T., *Nature* (1993) **363**, 603
3. Saito, R.*, et al.*, *Physical properties of carbon nanotubes*. Imperial College Press, London, (1998)
4. Chen, Z. H.*, et al.*, *Science* (2006) **311**, 1735
5. Avouris, P.*, et al.*, *Nat. Nanotechnol.* (2007) **2**, 605
6. Rosenblatt, S.*, et al.*, *Appl. Phys. Lett.* (2005) **87**, 153111
7. Chimot, N.*, et al.*, *Appl. Phys. Lett.* (2007) **91**, 153111
8. Sazonova, V.*, et al.*, *Nature* (2004) **431**, 284
9. Steele, G. A.*, et al.*, *Science* (2009) **325**, 1103
10. Lassagne, B.*, et al.*, *Science* (2009) **325**, 1107
11. Dresselhaus, M. S.*, et al.*, *Annu. Rev. Phys. Chem.* (2007) **58**, 719
12. Gabor, N. M.*, et al.*, *Science* (2009) **325**, 1367
13. Avouris, P.; Chen, J., *Materials Today* (2006) **9**, 46
14. Odom, T. W.*, et al.*, *J. Phys. Chem. B* (2000) **104**, 2794
15. Charlier, J. C.*, et al.*, *Rev. Mod. Phys.* (2007) **79**, 677
16. Kikkawa, J. M.; Awschalom, D. D., *Nature* (1999) **397**, 139
17. Loss, D.; DiVincenzo, D. P., *Phys. Rev. A* (1998) **57**, 120
18. Kuemmeth, F.*, Ilani, S., et al.*, *Nature* (2008) **452**, 448
19. Churchill, H. O. H.*, et al.*, *Nat. Phys.* (2009) **5**, 321
20. Churchill, H. O. H.*, et al.*, *Phys. Rev. Lett.* (2009) **102**, 166802
21. Kong, J.*, et al.*, *Nature* (1998) **395**, 878
22. Hafner, J. H.*, et al.*, *Chem. Phys. Lett.* (1998) **296**, 195
23. Liu, L.; Fan, S. S., *J. Am. Chem. Soc.* (2001) **123**, 11502
24. Liu, Z. A.*, et al.*, *J. Am. Chem. Soc.* (2008) **130**, 13540
25. Hanson, R.*, et al.*, *Rev. Mod. Phys.* (2007) **79**, 1217
26. Johnson, A. C.*, et al.*, *Nature* (2005) **435**, 925
27. Koppens, F. H. L.*, et al.*, *Science* (2005) **309**, 1346
28. Simon, F.*, et al.*, *Phys. Rev. Lett.* (2005) **95**, 017401
29. Taylor, J. M.*, et al.*, *Phys. Rev. Lett.* (2003) **90**, 206803
30. Bockrath, M.*, et al.*, *Nature* (1999) **397**, 598
31. Deshpande, V. V.; Bockrath, M., *Nat. Phys.* (2008) **4**, 314
32. McEuen, P. L.*, et al.*, *Phys. Rev. Lett.* (1999) **83**, 5098
33. Steele, G. A.*, et al.*, *Nat. Nanotechnol.* (2009) **4**, 363
34. Maultzsch, J.*, et al.*, *Phys. Rev. B* (2005) **72**, 205438
35. Lefebvre, J.*, et al.*, *Nano Lett.* (2006) **6**, 1603
36. Bachilo, S. M.*, et al.*, *Science* (2002) **298**, 2361
37. Sfeir, M. Y.*, et al.*, *Science* (2004) **306**, 1540
38. Chandra, B.*, et al.*, *Phys. Status Solidi B* (2006) **243**, 3359
39. Zhong, Z. H.*, et al.*, *Nat. Nanotechnol.* (2008) **3**, 201
40. Mintmire, J. W.*, et al.*, *J. Phys. Chem. Solids* (1993) **54**, 1835
41. Saito, R.*, et al.*, *Phys. Rev. B* (1992) **46**, 1804
42. Zhou, C. W.*, et al.*, *Phys. Rev. Lett.* (2000) **84**, 5604

43. Bushmaker, A. W., *et al.*, *Phys. Rev. Lett.* (2009) **103**, 067401
44. Ouyang, M., *et al.*, *Science* (2001) **292**, 702
45. Kane, C. L.; Mele, E. J., *Phys. Rev. Lett.* (1997) **78**, 1932
46. Minot, E. D., *et al.*, *Phys. Rev. Lett.* (2003) **90**,
47. Valavala, P. K., *et al.*, *Phys. Rev. B* (2008) **78**, 235430
48. Deshpande, V. V., *et al.*, *Science* (2009) **323**, 106
49. Minot, E. D., *et al.*, *Nature* (2004) **428**, 536
50. Moriyama, S., *et al.*, *Phys. Rev. Lett.* (2005) **94**, 186806
51. Cobden, D. H.; Nygard, J., *Phys. Rev. Lett.* (2002) **89**, 046803
52. Jarillo-Herrero, P., *et al.*, *Nature* (2005) **434**, 484
53. Jarillo-Herrero, P., *et al.*, *Nature* (2004) **429**, 389
54. Cao, J., *et al.*, *Nat. Mater.* (2005) **4**, 745
55. Lassagne, B., *et al.*, *Nano Lett.* (2008) **8**, 3735
56. Huttel, A. K., *et al.*, *Nano Lett.* (2009) **9**, 2547
57. Ando, T., *J. Phys. Soc. Jpn.* (2000) **69**, 1757
58. Wunsch, B., *Phys. Rev. B* (2009) **79**, 235408
59. Secchi, A.; Rontani, M., *Phys. Rev. B* (2009) **80**, 041404
60. Huertas-Hernando, D., *et al.*, *Phys. Rev. B* (2006) **74**, 155426
61. Chico, L., *et al.*, *Phys. Rev. Lett.* (2004) **93**, 176402
62. Izumida, W., *et al.*, *J. Phys. Soc. Jpn.* (2009) **78**, 074707
63. Jeong, J. S.; Lee, H. W., *Phys. Rev. B* (2009) **80**, 075409
64. Zhou, J., *et al.*, *Phys. Rev. B* (2009) **79**, 195427
65. Dora, B., *et al.*, *Phys. Rev. Lett.* (2008) **101**, 106408
66. Fang, T. F., *et al.*, *Phys. Rev. Lett.* (2008) **101**, 246805
67. Logan, D. E.; Galpin, M. R., *J. Chem. Phys.* (2009) **130**, 224503
68. Bulaev, D. V., *et al.*, *Phys. Rev. B* (2008) **77**, 235301
69. Galland, C.; Imamoglu, A., *Phys. Rev. Lett.* (2008) **101**, 157404
70. Hu, Y. M., *et al.*, *Phys. Rev. A* (2009) **80**, 022322
71. Sun, H., *et al.*, *Phys. Rev. B* (2009) **79**, 193404
72. Farmer, D. B.; Gordon, R. G., *Nano Lett.* (2006) **6**, 699
73. Graber, M. R., *et al.*, *Phys. Rev. B* (2006) **74**, 075427
74. Sapmaz, S., *et al.*, *Nano Lett* (2006) **6**, 1350
75. Biercuk, M. J., *et al.*, *Nano Lett.* (2005) **5**, 1267
76. Mason, N., *et al.*, *Science* (2004) **303**, 655
77. Buitelaar, M. R., *et al.*, *Phys. Rev. B* (2008) **77**, 245439
78. Jorgensen, H. I., *et al.*, *Nat. Phys.* (2008) **4**, 536
79. van der Wiel, W. G., *et al.*, *Rev. Mod. Phys.* (2003) **75**, 1
80. Ono, K., *et al.*, *Science* (2002) **297**, 1313
81. Palyi, A.; Burkard, G., *Phys. Rev. B* (2009) **80**, 201404(R)
82. Fischer, J., *et al.*, *Phys. Rev. B* (2009) **80**, 155401
83. Pfund, A., *et al.*, *Phys. Rev. Lett.* (2007) **99**, 036801
84. Petta, J. R., *et al.*, *Science* (2005) **309**, 2180
85. Reilly, D. J., *et al.*, *Appl. Phys. Lett.* (2007) **91**, 162101
86. Biercuk, M. J., *et al.*, *Phys. Rev. B* (2006) **73**, 201402

87. Gotz, G.*, et al.*, *Nano Lett.* (2008) **8**, 4039
88. Khaetskii, A. V.; Nazarov, Y. V., *Phys. Rev. B* (2001) **6412**,
89. Chan, V. C.*, et al.*, *J. Appl. Phys.* (2006) **100**, 106104
90. Hu, Y.*, et al.*, *Nat. Nanotechnol.* (2007) **2**, 622
91. Liu, H. W.*, et al.*, *Phys. Rev. B* (2008) **77**, 073310
92. Shaji, N.*, et al.*, *Nat. Phys.* (2008) **4**, 540

**Figures:**

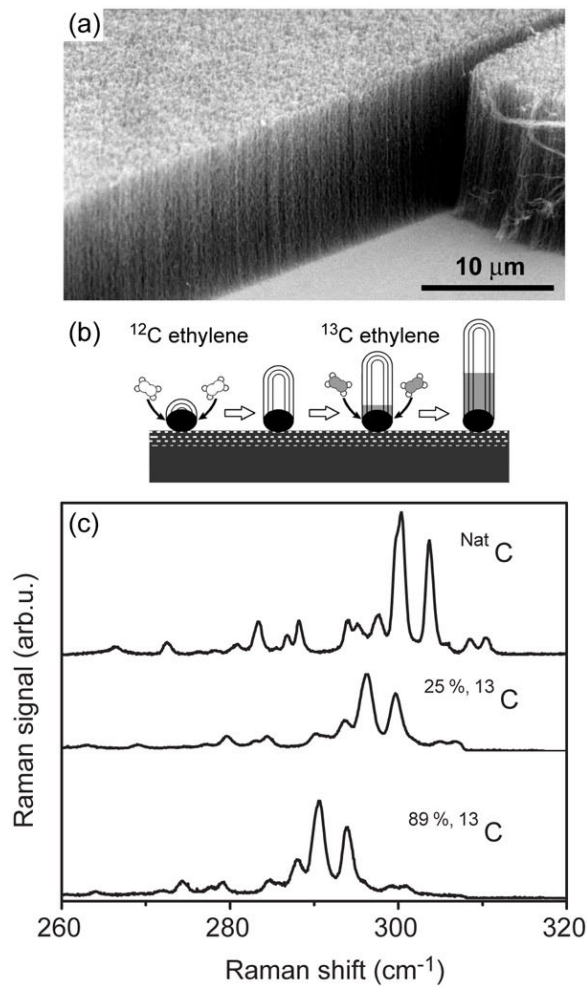

Fig. 1 (a) Scanning electron micrograph of a multi-wall NT array in which the isotopic composition changes along each nanotube from $^{12}C$ to $^{13}C$, top to bottom. (b) Isotopic labeling indicates growth by extrusion from catalyst particles fixed to the substrate. (c) Different techniques yield double-wall NTs in which the inner tube is predominantly $^{12}C$ (top trace) or $^{13}C$ (bottom trace). This shifts the frequency of the radial breathing mode of the inner nanotube, revealed here by Raman spectroscopy.
(Part (a,b) reprinted with permission from ref [23] L. Liu *et al.*, J. Am. Chem. Soc. (2001) **123**, 11502. Copyright 2001 American Chemical Society. Part (c) reprinted with permission from ref [28] F. Simon *et al.*, Phys. Rev. Lett. (2005) **95**, 017401. Copyright 2005 by the American Physical Society.)

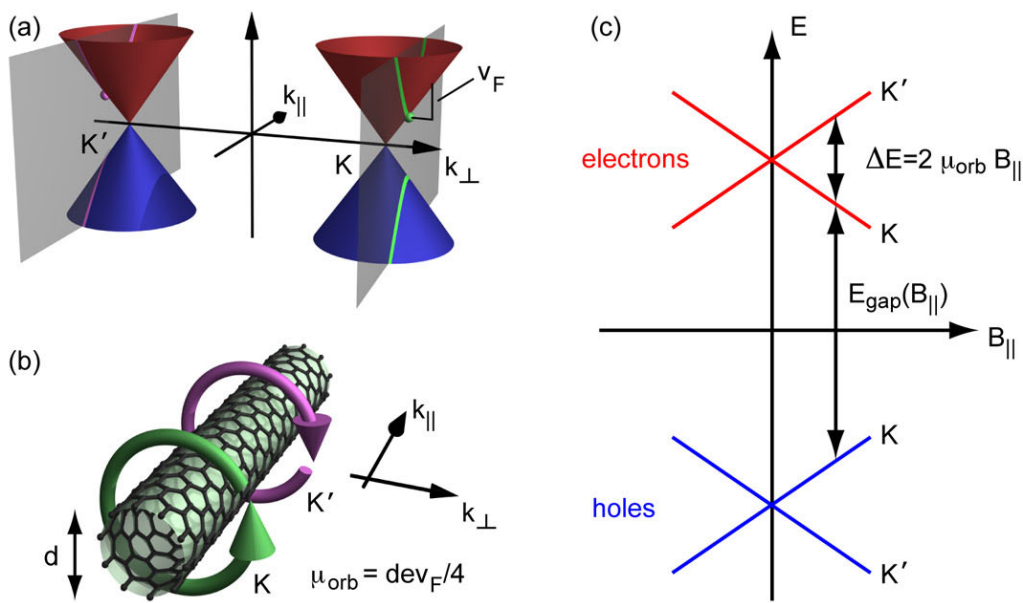

Fig. 2: Orbital electronic structure of NTs. (a) The quantization condition (gray planes) around the nanotube circumference results in four hyperbolic bands (green and purple) near the two Dirac points (K ,K') of graphene. (b) The lowest electron-like states in the K and K' valley are equal in energy (green and purple dots in panel a), and constitute a clockwise and counterclockwise persistent ring current. (c) The resulting large orbital magnetic moments $\mu_{orb}$ can be employed to lift the valley degeneracy or tune the band-gap $E_{gap}$ with an external magnetic field $B_{\parallel}$.

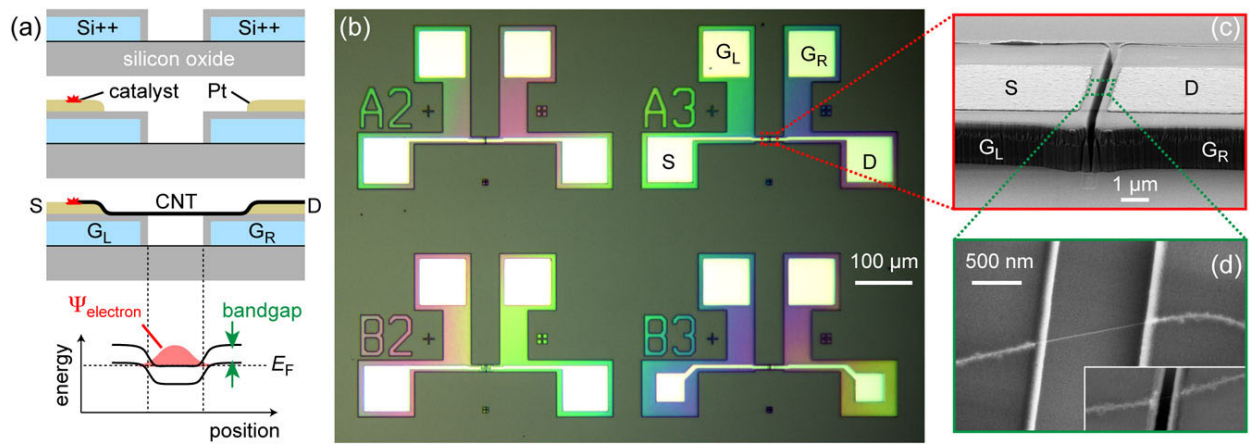

Fig. 3 (a) Schematic of the main fabrication steps of a suspended NT quantum dot. No processing is needed after CVD growth, resulting in clean NTs. (b) Optical image of the bonding pads for four devices before carbon nanotube growth. (c,d) Scanning electron micrographs of devices with 1 µm and 0.1 µm wide trenches. The nanotubes and oxide sidewalls appear bright in the top view (d).

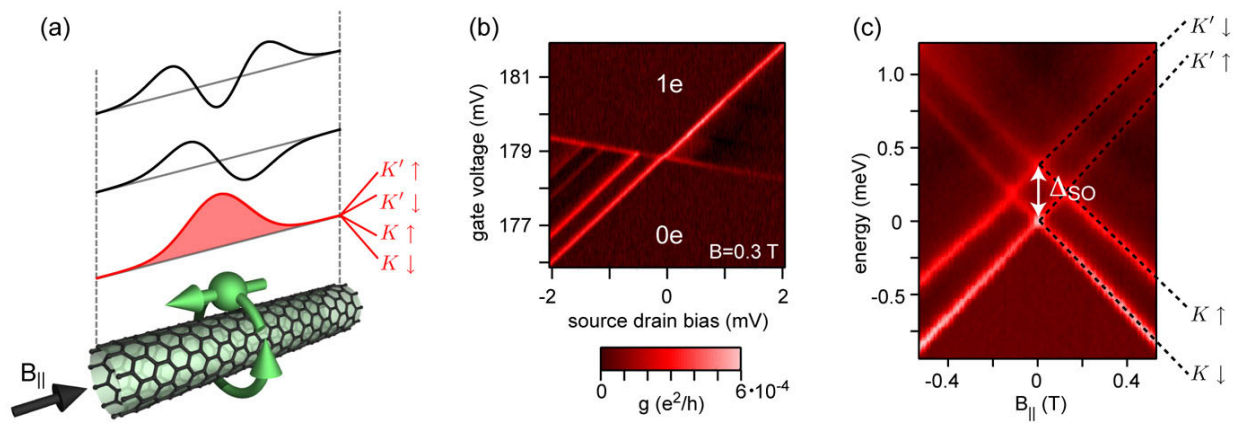

Fig. 4 (a) Understanding the four quantum states of the lowest electronic shell (red). (b) At sub-Kelvin temperatures and finite magnetic field the four combinations of clockwise/counterclockwise motion (K/K') and spin up/down (↑/↓) are clearly resolved in a suspended device using tunneling spectroscopy. (c) Their magnetic field dependence reveals a spin-orbit gap of $\Delta_{SO}$=0.37 meV at zero magnetic field[18].

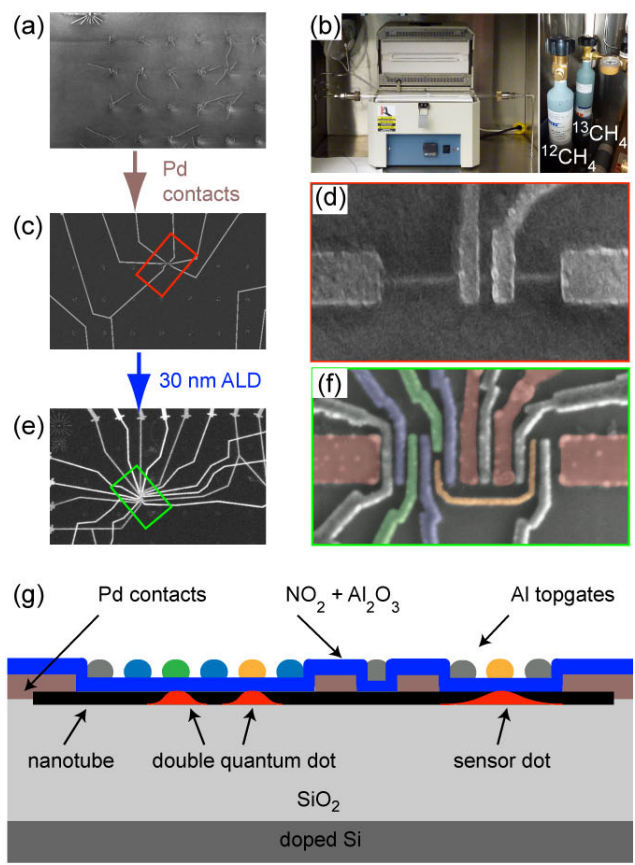

Fig. 5 (a) SWNTs of controlled $^{12}C/^{13}C$ composition are grown in a tabletop furnace from isotopically purified methane (b). Individual NTs are contacted by use of alignment marks (c,d) and gated (e) after atomic layer deposition (ALD) of a thin dielectric insulator. (f) The barrier gates (blue) and coupling wire (orange) allow the formation of a double quantum dot with integrated charge sensor on the same nanotube. (g) Schematic cross section of a finished device.

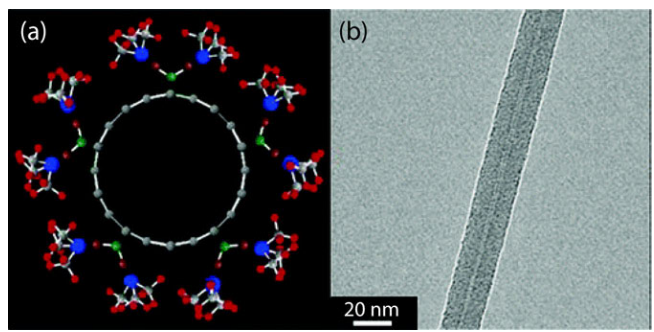

Fig. 6 (a) Schematic cross section of a $NO_2$ functionalized NT with one self-terminating monolayer of trimethylaluminum. (b) Transmission electron micrograph of a functionalized NT with a homogeneous coating of 10 nm aluminum oxide.
(Reprinted with permission from ref [72] D. B. Farmer *et al.*, Nano Lett. (2006) **6**, 699. Copyright 2006 American Chemical Society.)

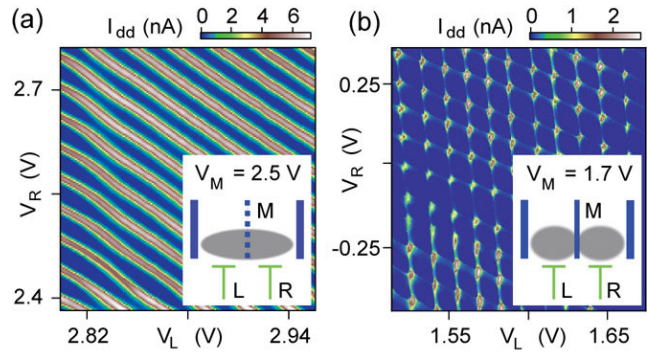

Fig. 7: Current $I_{dd}$ through the carbon nanotube as a function of right and left gate voltages. (a) Coulomb oscillations indicate a single quantum dot that is capacitively coupled to both the left (L) and right (R) plunger gate. (b) Increasing the middle barrier (M) transforms the quantum dot into a double quantum dot, and the conductance is suppressed except near triple points[19].

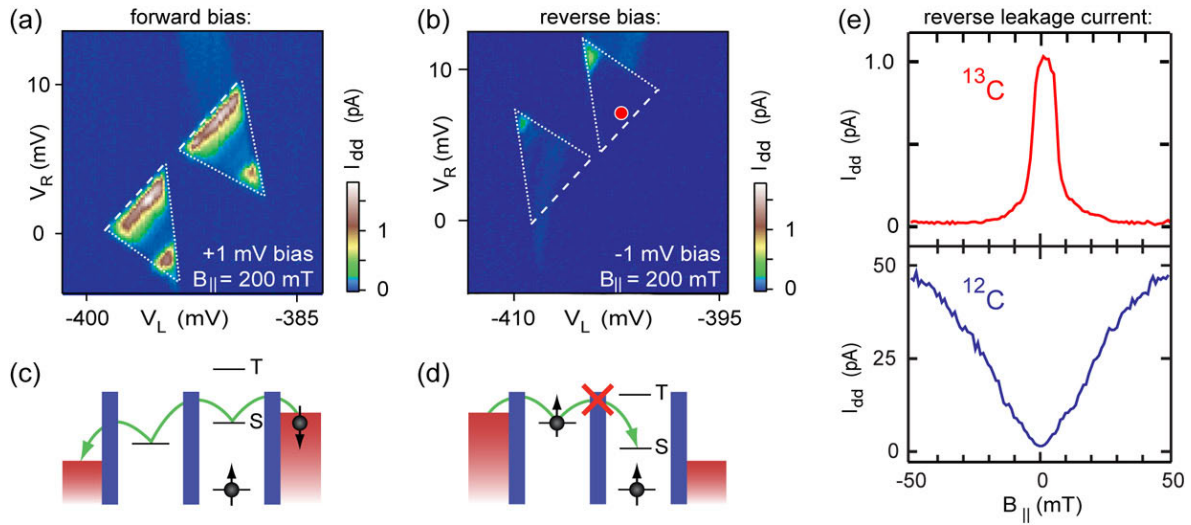

Fig. 8: Spin-blockade in a $^{13}C$ double quantum dot. (a) For positive bias current flow is observed near the base of the bias triangles (dashed line). (b) For negative bias current flow is strongly suppressed at finite magnetic field, indicating that spin selection rules prohibit interdot tunneling as schematically indicated in (d). (e) At zero magnetic field a small leakage current appears (top trace, measured near the red dot in panel b), indicating that electronic spins are efficiently flipped by nuclear spins. In contrast $^{12}C$ double quantum dots (lower trace) manifest a different behavior[19].

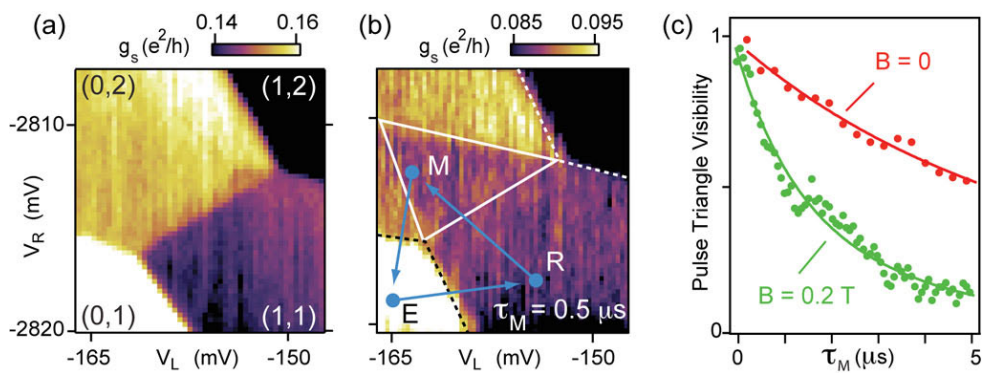

Figure 9: Pulsed-gate measurements of spin-relaxation using charge sensing. (a) The conductance of the charge sensor $g_S$ is a measure for the charge occupancy of the double quantum dot. (b) If the gate voltages are cycled between E, R and M faster than the spin-relaxation time, two separated electrons (1,1) may be prevented by their spin symmetry from tunneling into one dot (0,2). This results in a pulse triangle (white line) whose color sensitively depends on how much time $\tau_M$ is spent at point M. (c) The relaxation time is extracted from the dependence of the pulse triangle color on $\tau_M$ at various magnetic fields.

(Parts (a,b) reprinted with permission from ref [20] H. O. H. Churchill *et al.*, Phys. Rev. Lett. (2009) **102**, 166802. Copyright 2005 by the American Physical Society.)